\documentclass[10pt, conference, compsocconf]{IEEEtran}
\ifCLASSINFOpdf
\else
\fi
\usepackage[cmex10]{amsmath}
\usepackage[caption=false,font=footnotesize]{subfig}
\usepackage{amsmath,amssymb,epsfig}
\usepackage{float}
\usepackage{color}
\usepackage{graphicx}

\def\reals{\mathbb{R}}

\def\comp{\raise 1pt \hbox{$\scriptstyle\circ$}}

\def\upto{{\raise 1pt \hbox{$\scriptstyle \,\nearrow\,$}}}
\def\downto{{\raise 1pt \hbox{$\scriptstyle \,\searrow\,$}}}

\def\tos{\rightrightarrows}

\newtheorem{theorem}{Theorem}[subsection]

\newtheorem{definition}[theorem]{Definition}
\newtheorem{example}[theorem]{Example}
\newtheorem{remark}[theorem]{Remark}

\begin{document}
%
% paper title
% can use linebreaks \\ within to get better formatting as desired
\title{Semantic Content Filtering with Wikipedia and Ontologies}

% author names and affiliations
% use a multiple column layout for up to two different
% affiliations

\author{\IEEEauthorblockN{Pekka Malo, Pyry Siitari, Oskar Ahlgren, Jyrki Wallenius and Pekka Korhonen}
\IEEEauthorblockA{Department of Business Technology\\
Aalto University School of Economics\\
Helsinki Finland\\
E-mail for correspondence: pekka.malo@aalto.fi}
%\and
%\IEEEauthorblockN{Authors Name/s per 2nd Affiliation (Author)}
%\IEEEauthorblockA{line 1 (of Affiliation): dept. name of organization\\
%line 2: name of organization, acronyms acceptable\\
%line 3: City, Country\\
%line 4: Email: name@xyz.com}
}

% conference papers do not typically use \thanks and this command
% is locked out in conference mode. If really needed, such as for
% the acknowledgment of grants, issue a \IEEEoverridecommandlockouts
% after \documentclass

% for over three affiliations, or if they all won't fit within the width
% of the page, use this alternative format:
% 
%\author{\IEEEauthorblockN{Michael Shell\IEEEauthorrefmark{1},
%Homer Simpson\IEEEauthorrefmark{2},
%James Kirk\IEEEauthorrefmark{3}, 
%Montgomery Scott\IEEEauthorrefmark{3} and
%Eldon Tyrell\IEEEauthorrefmark{4}}
%\IEEEauthorblockA{\IEEEauthorrefmark{1}School of Electrical and Computer Engineering\\
%Georgia Institute of Technology,
%Atlanta, Georgia 30332--0250\\ Email: see http://www.michaelshell.org/contact.html}
%\IEEEauthorblockA{\IEEEauthorrefmark{2}Twentieth Century Fox, Springfield, USA\\
%Email: homer@thesimpsons.com}
%\IEEEauthorblockA{\IEEEauthorrefmark{3}Starfleet Academy, San Francisco, California 96678-2391\\
%Telephone: (800) 555--1212, Fax: (888) 555--1212}
%\IEEEauthorblockA{\IEEEauthorrefmark{4}Tyrell Inc., 123 Replicant Street, Los Angeles, California 90210--4321}}

% use for special paper notices
%\IEEEspecialpapernotice{(Invited Paper)}

% make the title area
\maketitle

\begin{abstract}
The use of domain knowledge is generally found to improve query efficiency in content filtering applications. In particular, tangible benefits have been achieved when using knowledge-based approaches within more specialized fields, such as medical free texts or legal documents. However, the problem is that sources of domain knowledge are time-consuming to build and equally costly to maintain. As a potential remedy, recent studies on Wikipedia suggest that this large body of socially constructed knowledge can be effectively harnessed to provide not only facts but also accurate information about semantic concept-similarities. This paper describes a framework for document filtering, where Wikipedia's concept-relatedness information is combined with a domain ontology to produce semantic content classifiers. The approach is evaluated using Reuters RCV1 corpus and TREC-11 filtering task definitions. In a comparative study, the approach shows robust performance and appears to outperform content classifiers based on Support Vector Machines (SVM) and C4.5 algorithm.
\end{abstract}


% For peer review papers, you can put extra information on the cover
% page as needed:
% \ifCLASSOPTIONpeerreview
% \begin{center} \bfseries EDICS Category: 3-BBND \end{center}
% \fi
%
% For peerreview papers, this IEEEtran command inserts a page break and
% creates the second title. It will be ignored for other modes.
\IEEEpeerreviewmaketitle

\section{Introduction}
% no \IEEEPARstart
Recently, ontologies have become a broadly accepted solution for integrating semantic knowledge into document modeling tasks. By using ontologies as a source of background knowledge, the IR expert systems have achieved increased contextual understanding and ability to do accurate conceptual indexing. Yet, these advantages are not gained without time-consuming ontology engineering. To reduce the costs of managing complicated knowledge models, there is an ongoing quest for alternative approaches. Therefore, an emerging trend is to consider the use of socially developed sources of semantic information, such as Wikipedia, to complement expensive domain ontologies; see Medelyan et al.~\cite{medelyan09}.

In this paper, we propose a new framework for document filtering, Wiki-SR\footnote{Wikipedia-based Semantic Rules}, where Wikipedia-based concept-relatedness information is integrated with a domain ontology to produce semantic document classifiers. In certain sense, this approach can be summarized as a rule-based filtering model where the filtering criteria are represented by "semantified" boolean queries. Here, the difference between an ordinary boolean query and a semantic filtering rule is that by using concept-relatedness information a semantic rule can match also such documents which do not explicitly feature the original query concepts. Thus, the main idea is essentially quite simple: each semantic rule provides a model for an "implicit expansion" of the original query by allowing the rule to match/accept also concepts which are not mentioned in the original query. The acceptance is done on the condition that the filtered document contains concepts which are strongly related to the concepts constituting the rule. 

The evaluation of the Wiki-SR framework was carried out using Reuters RCV1 corpus. The data set was chosen due to the relevance judgements and topic definitions supplied by the assessors of TREC-11 filtering track. As benchmarks,  we used the Support Vector Machines (SVM) and the decision-tree algorithm C4.5, which are well-known for their solid performance. Both algorithms were built using several different feature sets ranging from bag-of-words to Wikipedia- and ontology-based document models. As primary performance measures, we used F-score, precision, and recall. The overall result appeared very positive for the heuristic Wiki-SR model, which outperformed the benchmarks in terms of F-score by a fair margin. 

The rest of this paper is organized as follows. Section~\ref{sec:contributions} gives a short review of related work and summarizes the contributions of this paper. Section~\ref{sec:framework} provides an overview of the Wiki-SR framework. The components of the document model used by the semantic rules are introduced in Section~\ref{sec:conceptualization}. The notion of concept-relatedness measures and the definition of the Wiki-SR model are presented in Section~\ref{sec:wikisr}. An experiment based on the algorithm is given in Section~\ref{sec:experiment}. We conclude in Section~\ref{sec:conclusion}.

\section{Related work and contributions}\label{sec:contributions}

Today, Wikipedia is increasingly recognized as a valuable source of semantic knowledge for various natural language processing tasks; see Medelyan et al.~\cite{medelyan09} for a comprehensive review. As pioneering research in this field, we acknowledge the work done by Milne et al.~\cite{milne07,milne08}, Gabrilovich and Markovich~\cite{gabrilovich06,gabrilovich07}, Medelyan et al.~\cite{medelyan08a,medelyan08b}, Mihalcea and Csomai~\cite{mihalcea07}, and Strube and Ponzetto~\cite{strube06}, who have examined different ways of using Wikipedia to compute semantic relatedness between concepts and perform automated cross-referencing of documents. 

However, considering the large potential offered by Wikipedia, surprisingly little research has examined its use for document profiling, clustering and classification tasks. Perhaps, the best known papers, where Wikipedia has been used for information retrieval tasks, are the studies on query expansion by Gregorowicz and Kramer~\cite{gregorowicz06} and Milne et al.~\cite{milne07b}. Later, these have been followed by research on how pseudo-relevance feedback and explicit semantic analysis can be used to improve queries; see Li et al.~\cite{li07} and Egozi et al.~\cite{egozi08}. Among the latest studies are also the papers by Wang et al.~\cite{wang08a,wang08b} where semantic kernels are derived from Wikipedia to be used in SVM classifiers and co-clustering methods. 

In this paper, our main contribution to the existent literature is the introduction of Wikipedia-based semantic rules for document filtering. This technique capitalizes on the simplicity of ordinary boolean queries but improves it by performing an implicit expansion to take into account the actual semantic meanings of the concepts involved in the query. However, the idea in our Wiki-SR framework is quite different from what has become known as query expansion as considered by Milne et al.~\cite{milne07b}. Whereas explicit query expansion is commonly defined as addition of terms and phrases to the original query phrase to produce a more comprehensive and also more complex expression, we never add new terms to the original query. Instead, the synonyms and closely related concepts are taken into account implicitly through similarity measures in the first evaluation step of the semantic rule. Furthermore, although the Wiki-SR rules in certain sense build a semantic kernel to model concept-relatednesses, the system is closer to a semantic boolean query than a kernelised SVM-classifier.

The second contribution of this paper is concerned with the way of modelling document content. The model described in Section~\ref{sec:conceptualization} combines three different approaches: a Wikipedia, an ontology, and the classical bag-of-words content models. Here, the Wikipedia-based content model is further divided into sub-models representing general concepts and named-entities (NE) by using a Conditional Random Fields (CRF) classifier. The benefit is that this separation allows us to take into account the inherent differences in the narrowness of concept definitions. In addition to Wikipedia, we also utilise a small business ontology (BTO) to account for specialized economic concepts which are not equally well captured by Wikipedia. The BTO ontology also provides a well-defined hierarchy, which has proven to be effective in defining Wiki-SR rules.

\section{Wiki-SR framework}\label{sec:framework}

The Wiki-SR framework is an interactive content filtering system that combines the relevance statements supplied by the user with the concept-relatedness information in Wikipedia to produce semantic rules for identifying the documents that match the given topic. To summarize the steps involved in the filtering process, we split the overview of the framework into two parts: (1) the content modeling component; and (2) the Wikipedia-based semantic rule component.

\begin{figure}[h]
\centering
\includegraphics[scale=0.7]{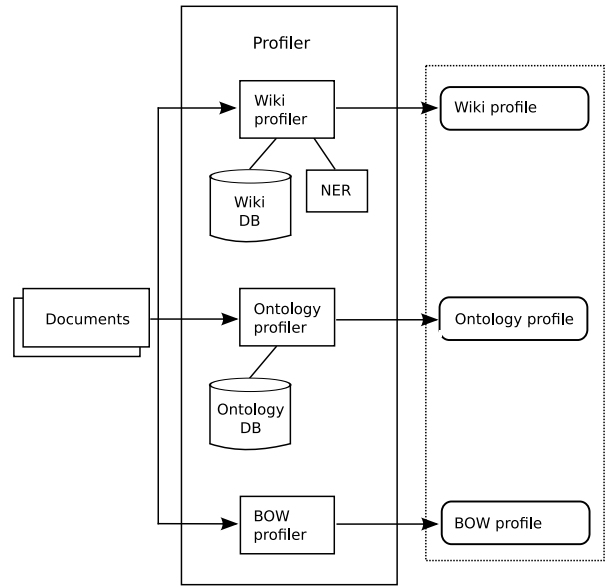} 
\caption{Content modeling component}\label{fig:contentmodel}
\end{figure}

The first component, {\it content modeling}, is shown in Figure~\ref{fig:contentmodel}. Once an incoming document has been preprocessed, the profile is constructed in three parts: a Wikipedia-content model (Section~\ref{sec:wikimodel}), an Ontology-content model (Section~\ref{sec:ontologymodel}), and the classical Bag of Words (BOW) representation. Together, these constitute the document model (Section~\ref{sec:docmodel}) used for filtering tasks. Although, there is overlap between the models, they tend to capture different aspects in the document, which makes them complementary. The resources used in profiling consists of the Wikipedia's link-structure (Wiki DB), the business term ontology BTO (Ontology DB), and named-entity recognizer (NER). For further details on document model, see Section~\ref{sec:conceptualization}.

\begin{figure}[h]
\centering
\includegraphics[scale=0.7]{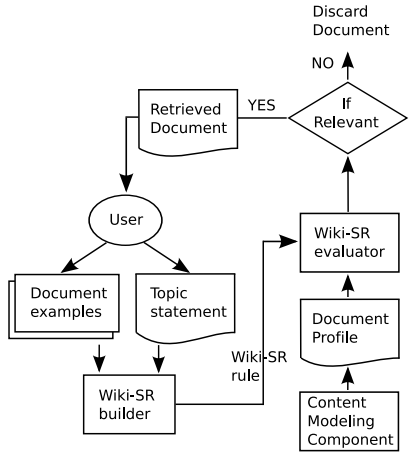} 
\caption{Semantic rule component}\label{fig:semanticrule}
\end{figure}

The second component, {\it Wikipedia-based semantic rule}, is described in Figure~\ref{fig:semanticrule}. The purpose is to represent the user's information needs as compositions of standard boolean queries, which are expressed in terms of Wikipedia and ontology concepts. The resulting rule is referred to as a Wikipedia-based semantic rule (Wiki-SR rule), which represents the topic of user's interest. 

\begin{figure}[h]
\centering
\includegraphics[scale=0.8]{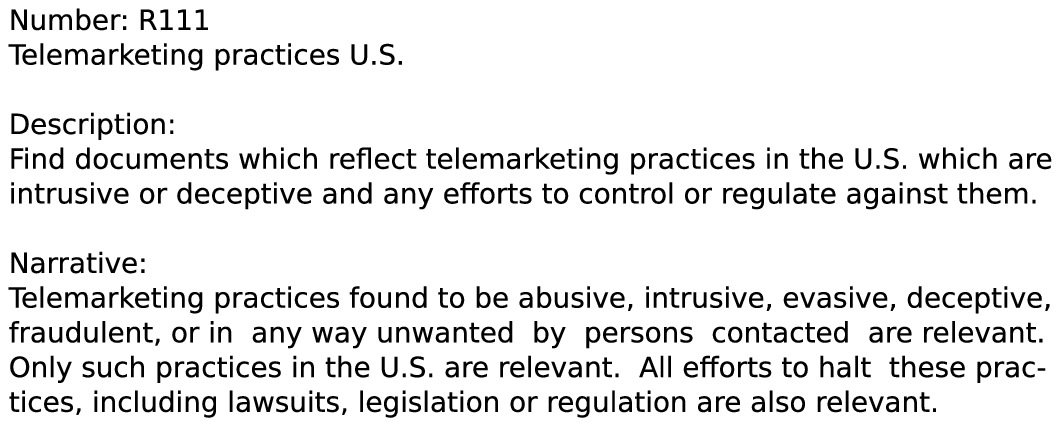} 
\caption{Example topic statement}\label{fig:topicstatement}
\end{figure}

In order to build the rules (Wiki-SR builder in Figure~\ref{fig:semanticrule}), the user is expected to supply a topic statement (see Figure~\ref{fig:topicstatement}) defining the central concepts, and a small set of relevant/irrelevant example documents that can be used as a training data for learning the Wiki-SR rule that best describes the given topic. Each topic statement stands for a single topic by providing a short textual description of the concepts which are relevant or irrelevant. For implementation of the Wiki-SR builder, see discussion in Section~\ref{sec:wikisrmodel}.

Once the semantic rule has been learned, it is given to an evaluator (Wiki-SR evaluator in Figure~\ref{fig:semanticrule}) which checks whether the incoming documents match the given rule based on their profiles. This is the stage where the main benefit of constructing the rules in terms of Wikipedia's concepts is realized. While checking the potential matches, the evaluator uses Wikipedia's concept-relatedness information to judge whether the query's concepts are present in the given document.

\section{Wikipedia-enhanced conceptualization}\label{sec:conceptualization}

Research on ontology-based knowledge models has been largely motivated by their ability to provide unique definitions for concepts, their relationships and properties, which together create a unified description of a given domain. However, the use of ontologies has been limited by the large engineering costs, which has stimulated increasing research on socially or automatically constructed knowledge-resources. In this section, we describe a hybrid document model, where Wikipedia is used in conjunction with a small ontology-based document model and the classical bag-of-words representation.

\subsection{Wikipedia-based content model}\label{sec:wikimodel}

Given a preprocessed document, we start building a document model by first detecting the Wikipedia-concepts\footnote{The notion Wikipedia-concept is used interchangeably with Wikipedia-article, because each article in Wikipedia represents a single topic/concept with a short title. In effect, this amounts to considering Wikipedia as a very large thesaurus consisting of the terms derived from the titles of all articles.} that best represent its contents. To identify the possible concepts, we use the machine learning approach described by Milne et al.~\cite{milne08} which is a refined version of the algorithm suggested by Medelyan et al.~\cite{medelyan08a}. There, the idea is to train a two-stage classifier, where the first classifier (disambiguator) recognizes terms that should be linked and the second classifier (link detector) decides where those terms should link to.
This automatic cross-referencing process is commonly known as {\it topic indexing} or {\it wikification}:
\begin{definition}[Wikifier]\label{def:wikifier}
Let $W$ be an instantiation of Wikipedia. Wikifier is defined as a set-valued mapping\footnote{In the rest of the paper, the notation $\tos$ is used to denote a set-valued mapping.}, $\text{Wikify}:\mathcal{D}\tos W$, from the collection of documents $\mathcal{D}$ to the set of all Wikipedia-concepts. That is, for a given document $d\in\mathcal{D}$, $\text{Wikify}(d)\subset W$ is a collection of links to Wikipedia-articles.
\end{definition}

In its current form, Wikifier makes no difference between general concepts and named-entities. However, as we acknowledge in Section~\ref{sec:wikisr}, there is a considerable difference in the specificity of a concept which is a named-entity, e.g. "Goldman Sachs", and a general concept, e.g. "Investment banking". For instance, to say that a certain document discusses Goldman Sachs practically requires that the bank's name is explicitly mentioned. But, to say that a document is about investment banking is considerably more relaxed; it is sufficient to find a collection of investment banking related concepts rather than the exact concept name to identify the document as relevant. Clearly, this should be taken into account when specifying the sensitivity of semantic classifiers to different concept types. Therefore, we train a named-entity recognizer to complement the Wikifier. 
\begin{definition}[Named-Entity Recognizer]
Let $\Sigma$ denote a language such that $W\subset\Sigma$. The named-entity recognizer is defined as a set-valued mapping, $\text{NER}:\mathcal{D}\tos\Sigma$, from documents to the collection of all n-grams in language $\Sigma$ which can be interpreted as named-entities. 
\end{definition}
\begin{remark}
The $\text{NER}$-mapping is implemented as the Conditional Random Fields (CRF) -based classifier proposed by Finkel et al~\cite{finkel05}. An advantage of this model is that the system is able to augment non-local information, which allows for long-distance dependency models and enforcing of label consistency.
\end{remark}

Finally, having obtained both the set of Wikipedia concepts and the set of named-entities, we construct the Wiki-content model as a combination of the general Wikipedia-concepts and Wikipedia named-entities.

\begin{definition}[Wiki-content model]
Let $W$ be an instantiation of Wikipedia. The Wiki-content model is defined as the set-valued mapping, $\Lambda_W:\mathcal{D}\tos W$, which is given by the union, $\Lambda_W(d)=\Lambda_N(d)\cup\Lambda_G(d)$, where 
\begin{itemize}
\item[(i)] the model for Wikipedia named-entities, $\Lambda_N:\mathcal{D}\tos W$, is given by $\Lambda_N(d):=\text{NER}(d)\cap\text{Wikify}(d)$; and 
\item[(ii)] the model for general Wikipedia concepts, $\Lambda_G:\mathcal{D}\tos W$, $\Lambda_G(d):=\text{Wikify}(d)\setminus\Lambda_N(d)$, corresponds to the remainder of concepts identified by Wikifier.
\end{itemize}
\end{definition}

\subsection{Ontology-based content model}\label{sec:ontologymodel}

The ontology model considered in this paper is derived from the Business Term Ontology (BTO) proposed by Malo and Siitari~\cite{malo10}. The primary purpose of the BTO ontology is to provide the system with a solid taxonomy of business domain concepts, and allow explicit expression of generality vs. specificity of concepts through subclassing relation. 

The BTO-ontology model is built using the  RDFS extension proposed by Suchanek et al.~\cite{suchanek08}, where an ontology $\mathcal{O}$ is defined as an injective mapping from a finite set of fact-identifiers to fact-triplets. This definition allows a very general description of an ontology as a graph, where the nodes may be either entities (e.g. concepts such as OptionContract, PutOption, CallOption), relations (e.g. subClassOf, hasWikiPage) or fact-identifiers. The basic element in the BTO model is thus an entity which may refer to any abstract or concrete thing. Throughout, we also assume that the entities are discernible and we can tell whether two entities are the same.

Following the notations introduced in the previous section, we can now define the BTO-content model as a simple set-valued mapping:
\begin{definition}[BTO-content model]
Let $\mathcal{O}_{\text{BTO}}$ denote the current instantiation of the BTO-ontology, and $\mathcal{C}\subset\Sigma$ be the set of ontology concepts expressed in language $\Sigma$. The ontology-based content profiler is defined as a set-valued mapping, $\Lambda_{\mathcal{O}}:\mathcal{D}\tos\mathcal{C}$, from the collection of document $\mathcal{D}$ to the set of BTO-concepts. Then the BTO-content model for document $d\in\mathcal{D}$ is given by the set of concepts $\Lambda_{\mathcal{O}}(d)\subset\mathcal{C}$ produced by the profiler.
\end{definition}

\subsection{Document model}\label{sec:docmodel}

The full document model is then obtained as a combination of the Wikipedia and ontology based content models, which are augmented with the classical Bag of Words (BOW) representation.

\begin{definition}[Document model]\label{docmodel}
Let $\Lambda_W(d)$ and $\Lambda_{\mathcal{O}}(d)$ denote the Wikipedia and BTO content models for an active document $d\in\mathcal{D}$, respectively, and let $\Lambda_{\Sigma}:\mathcal{D}\tos\Sigma$, $\Lambda_{\Sigma}(d)\subset\Sigma$, be the bag of words (BOW) representation of document $d$ for the given language $\Sigma$. The document model is given by 
$$
\Lambda(d):=\left(\Lambda_W(d),\Lambda_{\mathcal{O}}(d),\Lambda_{\Sigma}(d)\right)
$$
which is interpreted as a sparse vector in $\reals^N$, where $N$ is the joint cardinality of Wikipedia $W$, BTO-ontology's concept-set $\mathcal{C}$, and the language $\Sigma$.
\end{definition}

This choice of document model leads to a standard vector-space representation, where the presence/absence of a Wikipedia-concept, BTO-concept or a word is indicated by ones and zeros. Although, the above form could be easily changed to support some weighting scheme, such as Tf-Idf, we have left them as a question for further research. For the purpose of the current experiment our primary interest is in the benefits obtained from the use of Wikipedia-based relatedness measures in detection of relevant documents.

\section{Semantic filtering with Wiki-SR}\label{sec:wikisr}

To illustrate the notion of semantic filtering in Wiki-SR, let's consider the sample topic statement (see Figure~\ref{fig:topicstatement}), where the goal is to filter documents reporting telemarketing abuses in U.S. Now, by reading the statement, one could come up with a boolean query to represent the topic; e.g. ("U.S." $\star$ "telemarketing" $\star$ ("fraudulent" $+$ "legislation" $+$ "regulation")\footnote{AND ($\star$), OR ($+$)}. Clearly, this kind of a word-based boolean expression looks reasonable on the surface and could itself be used directly, but it is unlikely to yield good recall or precision. The problem is that there are large amounts of synonymous or strongly related concepts which could appear in the documents instead of the original ones. For example, if a document has words \{"telesales", "Boston", "crime"\}, then it is likely to be a match because the words are almost synonyms or very strongly related to the query - even though none of these words appear in the original query. Thus, to enrich the original query with semantic knowledge, we need to find a way to measure the relatedness between any arbitrary pair of concepts and incorporate this idea into the evaluation of the query. For this purpose, we have designed the Wiki-SR model, which performs an implicit expansion of the query by using Wikipedia's relatedness information.

In order to clarify more closely what Wiki-SR rules are, how they are constructed, and evaluated in practice, the section is divided into the following parts. In the first part (Section~\ref{sec:relatedness}), we discuss how Wikipedia can be used to compute semantic relatedness between any pair of concepts. In particular, we consider how the existing measures can be adapted for usage in Wiki-SR rules. In the second part (Section~\ref{sec:wikisrmodel}), we present the formal definition of Wiki-SR model and discuss how it is constructed using the topic statement and the set of example documents supplied by the user. Finally, we describe the steps involved in evaluation of the Wiki-SR rules to determine whether a particular document matches the rule or not.

\subsection{Measuring semantic relatedness}\label{sec:relatedness}

Although approaches to measuring conceptual relatedness based on corpora or WordNet have been around already quite long, the use of Wikipedia as a source of background knowledge is a relatively new idea. The first step in this direction was taken by Strube and Ponzetto~\cite{strube06}, who proposed their WikiRelate-technique that modified existing measures to better work with Wikipedia. This was soon followed by the paper of Gabrilovich and Markovitch~\cite{gabrilovich07}, who suggested explicit semantic analysis (ESA) to define a highly accurate similarity measure using the full text of all Wikipedia articles. The most recent proposal is, however, the Wikipedia Link-based Measure (WLM) proposed by Milne et al.~\cite{milne07,milne08}, where only the internal link structure of Wikipedia is used to define relatedness. The approach is known to be computationally very cheap and has still achieved relatively high correlation with humans, which is why we have adopted it as a basis for the document-concept similarity measure used in this paper. Below, we describe how semantic relatedness information of Wikipedia-links can be incorporated into filtering rules.

Commonly, a semantic relatedness measure is defined between two concepts. However, from our application's perspective it is perhaps more interesting to ask: How strongly is the given concept related to the document at hands? Or how likely is it for the given concept to appear in the document? The idea of Milne et al.~\cite{milne07,milne08} was to construct a low-cost measure for semantic relatedness using only the hyperlink structure of Wikipedia rather than its category hierarchy or text content. The relatedness measure essentially corresponds to the Normalized Google Distance inspired by Cilibrasi and Vitanyi~\cite{cilibrasi07}:
\begin{definition}[Link-relatedness]
Let $w_1,w_2\in W$ be Wikipedia-concepts, and let $W_1, W_2 \subset W$ denote the sets of all articles that link to $w_1$ and $w_2$, respectively. The link structure -based concept-relatedness measure, $\text{link-rel}:W\times W \to [0,1]$ , is then given by
$$
\text{link-rel}(w_1,w_2)=\frac{\log{(\max{|W_1|,|W_2|})}-\log{(|W_1\cap W_2|)}}{\log{(|W|)}-\log{(\min{(|W_1|,|W_2|)})}}.
$$
\end{definition}

Because $\text{link-rel}$ is defined only for uniquely identified Wikipedia-concepts, we need to extend the definition slightly to allow relatedness calculation for any pair of n-grams. That is, many words which are recognized as redirects or anchors are not counted into the set of Wikipedia-concepts $W\subset\Sigma$. Therefore, we consider the following extension of $\text{link-rel}$ from $W$ to $\Sigma$.

\begin{definition}[Between terms -relatedness]
Let $s_1,s_2\in\Sigma$ be two terms. The Wikipedia-based term-relatedness measure, $\text{rel}:\Sigma\times\Sigma \to [0,1]$, is defined as
\begin{eqnarray*}
\text{rel}(s_1,s_2)&=&\max\{\text{link-rel}(w_1,w_2)\} \ : \ w_1\in \text{Senses}(s_1), \\
&&w_2\in \text{Senses}(s_2) \},
\end{eqnarray*}
where 
\begin{eqnarray*}
\text{Senses}(s_i)&=&\{w\in W : \text{$s_i$ is redirect, anchor, or title}\\ 
 && \text{of $w$} \}.
\end{eqnarray*} If $s_i$ is uniquely identified, i.e. $s_i\in W$, then $|Senses(s_i)|=1$. Therefore, if the terms are uniquely identified as Wikipedia-concepts, i.e. $s_1,s_2\in W$, then $\text{rel}(s_1,s_2)=\text{link-rel}(s_1,s_2)$. Also, if $\text{Senses}(s_i)\equiv\emptyset$ for some $s_i\in\Sigma$, then the term is not recognized by Wikipedia, and we have $\text{rel}(s_i,s_j)=0$ for every $s_j\in\Sigma$.
\end{definition}

Finally, recalling that we wanted a measure between a document and a concept, we can now use the above extension to introduce the following simple definition for Wikipedia-based relatedness measure:
\begin{definition}[Document-term relatedness]\label{def:rel}
Let $s\in\Sigma$ and $d\in\mathcal{D}$. The Wikipedia-based document-term -relatedness measure, $\text{d-rel}:\Sigma\times\mathcal{D}\to[0,1]$, is given by
$$
\text{d-rel}(s,d)=\max\{\text{rel}(s,\bar{s}) \ : \ \bar{s}\in\Lambda(d)\}
$$
where $\Lambda(d)$ is the document model~\eqref{docmodel}. By taking the maximum over $\Lambda(d)$ we allow $\bar{s}$ to be either a Wikipedia-concept, Ontology-concept or word in BOW.
\end{definition}
\begin{remark}
If $\bar{s}\in\Lambda_{\mathcal{O}}(d)$, then the relatedness is calculated with respect to the Wikipedia-page attached to the ontology concept through hasWikiPage-relation, i.e. we treat the concept as a Wikipedia-article. If the concept does not have Wikipedia-page defined, then it is treated as any n-gram. 
\end{remark}

The use of maximum, rather than sum-based operator such as average, in d-rel is a deliberate choice. Since this relatedness measure is intended to be used in filtering rules, we do not want to allow sum-operations to mask the presence of those concepts in a document which are not related to its central story.

\subsection{Wiki-SR model}\label{sec:wikisrmodel}

Having introduced the semantic relatedness measure, d-rel, we can now provide a more detailed explanation to semantic filtering rules. Following our earlier discussion in Section~\ref{sec:framework}, we decompose the definition of a semantic rule into two parts (see Figure~\ref{fig:semanticrule}): (1) the rule-builder which is responsible for learning the underlying query expression; and (2) the rule-evaluator which uses Wikipedia's concept-relatedness information to perform an implicit expansion of the query to account for strongly related concepts. 

\subsubsection{Rule-builder}

Let $V\subset\Sigma$ be the set of available ontology and Wikipedia concepts, and let $Q$ denote the space of all possible boolean query expressions that can be formulated using the concepts in $V$ and the boolean operators AND $(\star)$, OR $(+)$, and NOT $(\neg)$. 

Now, assuming that the user has provided a topic statement $t\in\mathcal{D}$ and a small training collection of relevant/irrelevant document examples $D_t\subset\mathcal{D}$, the rule builder is defined as a mapping from the user-inputs to the query space, i.e. 
$$
B: (t,D_t)\mapsto q\in Q,
$$ 
where $q$ can contain only those concepts which appear in the topic statement. That is, if $V_q=\{v_1,\dots,v_n\}\subset V$ is the set of concepts included in $q$, then the concepts must be such that $V_q\subset\Lambda_{W}(t)\cup\Lambda_{\mathcal{O}}(t)$.

\begin{example}\label{ex:builder}
Suppose that the user has provided the topic statement $t$ shown in Figure~\ref{fig:topicstatement}. After wikification of the document, we have identified a collection of concepts \{UnitedStates, Espionage, Fraud, Legislation, Regulation\}. Given the set of example documents $D_t$ by the user, the rule builder could produce a rule $B(t,D_t)=UnitedStates\star Espionage \star(Fraud+Legislation+Regulation)$.
\end{example}

The builder mapping $B$ is implemented by using the genetic programming (GP) technique proposed by Malo et al.~\cite{malo10b} that extends the Inductive Query By Example (IQBE) paradigm of Smith and Smith~\cite{smith97} and Chen et al.~\cite{chenshanka98}. There, the idea is to use the relevance information collected from the user as fitness cases to find the query expression that best separates relevant from irrelevant document examples. The learning process is driven by the evolutionary pressure that guarantees that only the fittest individuals among all potential query candidates survive. In this paper, we used F-score as the fitness function to find a reasonable balance between precision and recall. 

The reason, why IQBE-based query builders seem to be rarely used, is perhaps best explained by the tendency of GP to produce overfitted queries. The risk of overfitting is high, in particular, when the training sets are small and when the number of concepts (or literals/terminals in GP) is large. For these reasons, we had restricted the concept set to the ones that are detected from the topic statement. Thus, the version of GP used in this paper is a special case of the more advanced algorithm proposed by Malo et al.~\cite{malo10b}, where it is shown that the query learning can be generalized also to more realistic cases where predefined topic statements are not available. For further details on the use of GP-learning, see Koza~\cite{koza92}. To find more information on the ways how GP can be modified for learning Wikipedia-based queries, see the forthcoming paper by Malo et al.~\cite{malo10b}. 

\subsubsection{Rule-evaluator}

The rule-evaluator in Wiki-SR provides a matching subsystem for determining whether a given document matches the currently active semantic rule. Now, assuming that the user's topic definition has been transformed by the rule-builder component to a query $q\in Q$, the evaluator is specified as a binary-valued mapping,
$$
E:Q\times\mathcal{D}\to\{0,1\}.
$$
which operates in two steps: (i) concept-evaluation step;  and (ii) expression-evaluation step. 

To outline the procedure, let's suppose that the query expression has form $q=v_1 r v_2 r \cdots r v_k$, where $\{v_1,\dots,v_k\}\subset V$ and each $r$ could be replaced by any of the boolean operators. Then the evaluation steps can be defined as follows:

(i) \textit{Concept-evaluation step}:  The purpose of the concept-evaluation step is to determine whether the query concepts are present in the active document $d$ - either directly or indirectly. The task is accomplished by a specific concept-evaluator function that is applied in turn to every concept in the query. 

The concept-evaluator's decision rule is carried out in three parts: (a) First, it tries to look whether the query concepts are directly featured in the document. (b) If no match is found, the rule then searches for related concepts or words, which would strongly predict the presence of the given concepts. If the d-rel based sensitivity threshold is exceeded, then the rule decides that the given concept is present in the document. (c) Finally, if these two steps fail, it is concluded that the given concept is not present. 

To formalize this idea, we define the concept evaluator as function $\delta:V\times\mathcal{D}\to\{0,1\}$,
\[
\delta(v,d)=
\begin{cases}
1 & \text{if $v\in \Lambda_W(d)\cup\Lambda_{\mathcal{O}}(d)$},\\
1 & \text{if $v\in \text{Rel}(d)$}, \\
0 & \text{otherwise}
\end{cases}
\]
where $\text{Rel}(d)=\{v\in V : \text{d-rel}(v,d)>c_{\text{rel}}(v)\}$, and $c_{\text{rel}}>0$ is a threshold function controlling the acceptance sensitivity by relatedness criteria. The threshold for d-rel depends on the type of concept, i.e. whether it is a named-entity, general Wikipedia-article, or BTO-concept,
$$
c_{\text{rel}}(v)=
\begin{cases}
c_1 & \text{if $v$ is a named-entity, i.e. $v\in\Lambda_N(d)$}, \\
c_2 & \text{otherwise}. 
\end{cases}
$$
Each sensitivity threshold is chosen based on training data. The purpose of the distinction between named-entities and general concepts is to allow stricter thresholds for named-entities which have by default narrower definitions than general concepts.

\begin{example}
Let's continue our Example~\ref{ex:builder}, where the query rule produced by the builder was $q=UnitedStates\star Espionage \star(Fraud+Legislation+Regulation)$. Now, suppose that a new document $d$ features concepts $\{TradeSecret, China, Lawyer\}$. Then the concept-evaluator's task is to find $\delta(v,d)$ for every concept $v$ in $q$. In this particular case, we would obtain that $\delta(Espionage,d)=1, \delta(Legislation,d)=1$ because $Espionage$, $Lawyer\in Rel(d)$. However, $\delta(UnitedStates,d)=0$ because none of the concepts in $d$ is strongly related to the U.S.
\end{example}

(ii) \textit{Expression-evaluation step}: In Wiki-SR framework, it is the concept-evaluator function $\delta$ which does most of the work. Once the variables in the query $q$ have been evaluated, the expression-evalution step amounts to replacing the original query variables $\{v_1,\dots,v_k\}$ with the values given by the concept-evaluator $\{\delta(v_1,d),\dots,\delta(v_k,d)\}$, i.e. we obtain that
$$
E(q,d):=\delta(v_1,d)r\delta(v_2,d)r\cdots r\delta(v_k,d).
$$ 
Thus, the value of this final expression can then be used to decide whether the given document is relevant or not. For instance, in the case of the previous example we would find that the particular document is not relevant, because $\delta(UnitedStates,d)$ was 0.

\section{Experiment}\label{sec:experiment}

\subsection{Data}

The evaluation of Wiki-SR framework is based on Reuters RCV1 corpus\footnote{Reuters corpus volume 1, http://about.reuters.com\/researchandstandards\/corpus} using TREC-11 topic statements\footnote{TREC 2002 Filtering Track Collections, http://trec.nist.gov\/data}. The corpus contains about 800 000 news stories from years 1996-1997. Following TREC-11 instructions, the data set is partitioned into a training set (items dated between 1996-08-20 to 1996-09-30) and a test set (remainder of the collection). The training and test set are further divided into 100 topic-specific subsets, which are augmented with the relevance judgements made by the assessors of TREC-11. In this paper, only the initial training data is used, while the relevance statements available for adaptive learning are not utilized. 

\subsection{System}

The system used in the experiment was implemented using Java software on top of the GATE~\cite{bontcheva04} platform, which provides tools for standard document preprocessing tasks. The Wikipedia-based content model was built using the WikipediaMiner published by Milne et al.~\cite{milne09}, which was suitably modified and integrated into our framework. The named-entity recognition task was carried out using a Conditional Random Field (CRF) classifier proposed by Finkel et al.~\cite{finkel05}. For other classification tasks, we used Weka~\cite{hall09} through Java-ML~\cite{abeel09} package. The manual ontology editing was done in Prot\'eg\'e\footnote{http://protege.stanford.edu} framework. All automated ontology engineering tasks were done using Sesame\footnote{http://www.openrdf.org} with MySQL-repository.  

\subsection{Results}

In the experiment, the performance of Wiki-SR framework is compared against Support Vector Machines (SVM) and the decision-tree classifier C4.5. The primary performance measures used for comparison are F-score, precision, recall, and accuracy. 
\begin{figure*}
\centering
\subfloat[F-score]{\includegraphics[angle=-90,width=3in]{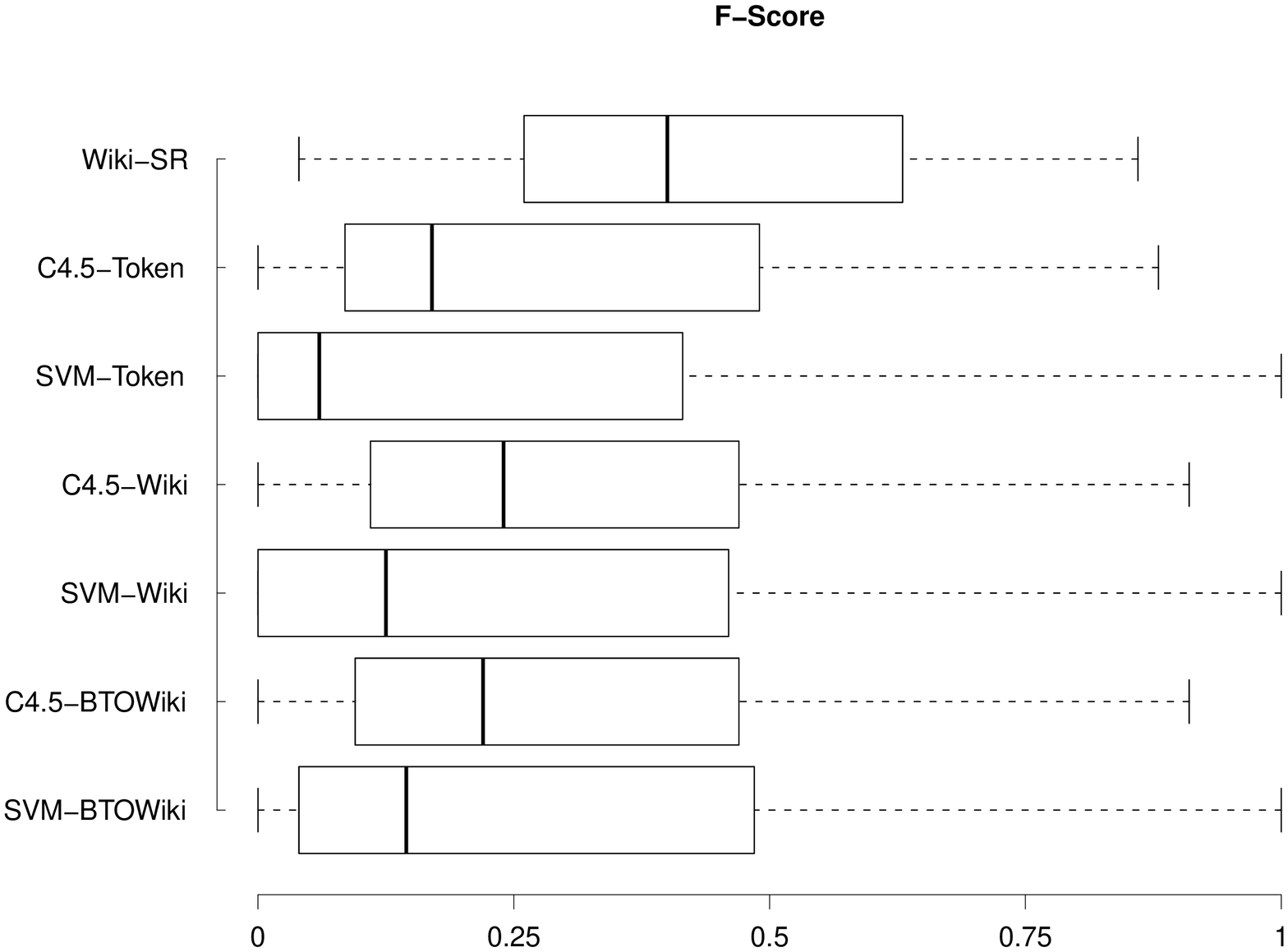}
\label{fig_first_case}}
\hfil
\subfloat[Recall]{\includegraphics[angle=-90,width=3in]{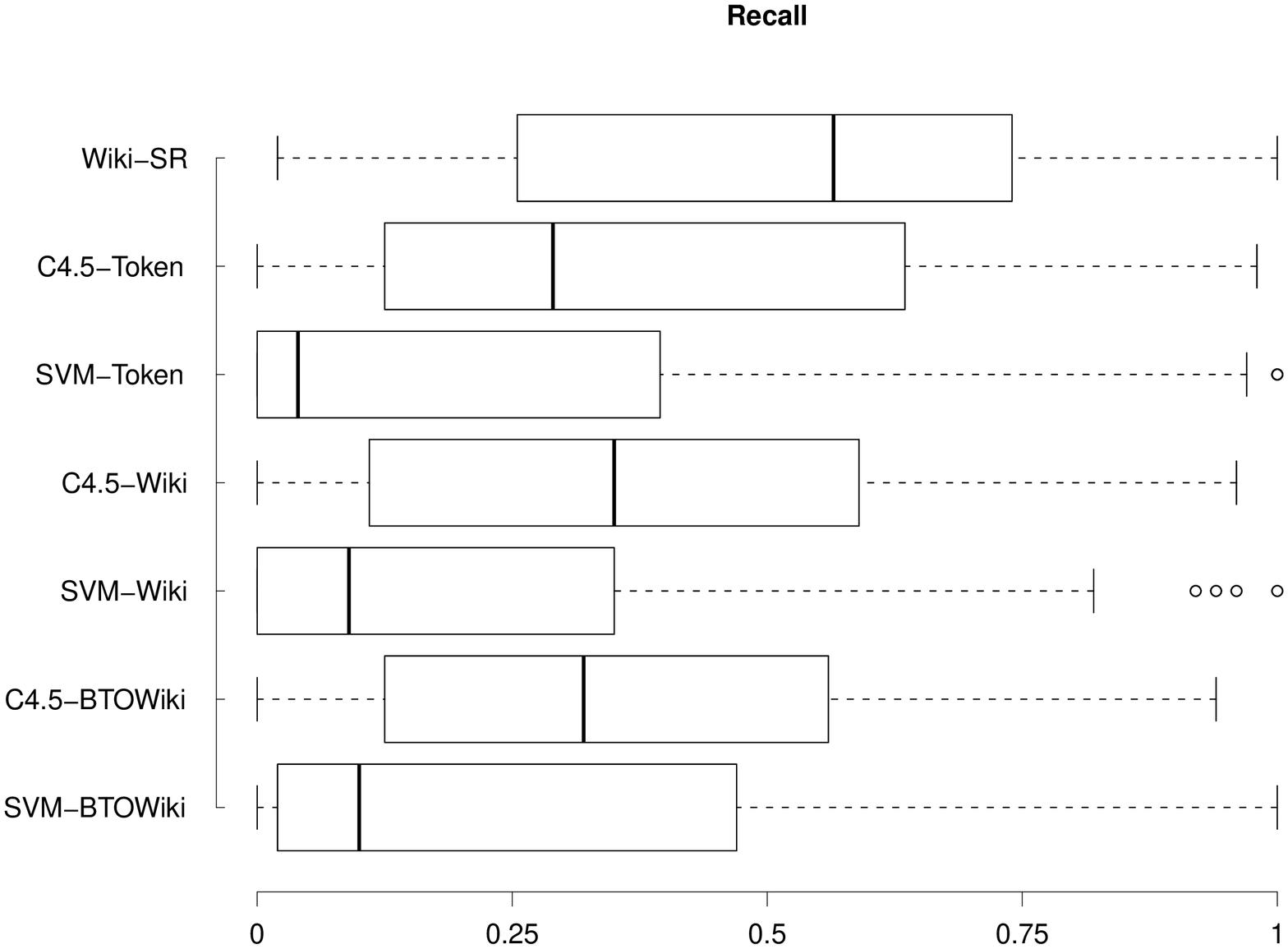}
\label{fig_second_case}}
\caption{Results for 100 TREC-11 topics}
\label{fig:comparisons}
\end{figure*}

The comparison was carried out as follows. The system started with the given collection of 100 topics and a set of training documents for each topic, where the documents had been pre-assigned as relevant or irrelevant by assessors of TREC-11. The task was then to train the classifiers using the information in the training-samples and the initial topic-statements. Here, each topic was considered separately and no cross-topic learning was allowed. 

The construction of Wiki-SR classifiers was implemented in two steps. First, in order to obtain the boolean query statements expressed in terms of Wikipedia and ontology concepts, the topic-statements were profiled and the obtained concepts were used to build query rules. Then, the sensitivity thresholds required by the relatedness-based acceptance criteria were optimised using the training samples. 

In similar fashion, the benchmark classifiers were optimised using only the training data. However, none of the benchmark classifiers used information in the original topic-statements. As feature sets, three different document models were considered: a bag-of-words profile (tokens), a Wikipedia profile (Wiki), and a profile where Wikipedia concepts are augmented with ontology concepts (BTO Wiki). 

Table~\ref{tab:comparison} reports performance measures for models Wiki-SR, LibSVM and C4.5. In order to take the varying quality of the different topics into account, the results are further divided into three subtables based on the ratio of positive and negative examples in the training sample, i.e. $tr=\text{\#negative examples} / \text{\#positive examples}$. Panel A gives results for all topics, Panel B for topics with low ratio ($tr\leq 5$), and Panel C for topics with high ratio ($tr> 5$).

\begin{table}[h]
\caption{Model comparison}\label{tab:comparison}
\begin{scriptsize}
\centering

\begin{tabular} {| l l || c |c | c | c |} 
\multicolumn{6}{l}{Panel A: Results for all topics} \\
\hline \hline 
{\em Model} & {\em Profile} & { F-Score } & { Accuracy } & { Precision } & { Recall } \\ \hline \hline{\em C4.5 } & {\em Tokens } & { 0.28 } & { 0.8 } & { 0.28 } & { 0.37 } \\
{\em } & {\em Wiki } & { 0.31 } & { 0.84 } & { 0.35 } & { 0.37 } \\
{\em } & {\em BTO Wiki } & { 0.30 } & { 0.83 } & { 0.32 } & { 0.36 } \\ \hline
{\em LibSVM } & {\em Tokens } & { 0.22 } & { 0.89 } & { 0.58 } & { 0.21 } \\
{\em } & {\em Wiki } & { 0.25 } & { 0.89 } & { 0.56 } & { 0.23 } \\
{\em } & {\em BTO Wiki } & { 0.27 } & { 0.88 } & { 0.54 } & { 0.25 } \\ \hline
{\em Wiki-SR} & {\em } & { 0.44 } & { 0.84 }& { 0.47 } & { 0.54 }  \\
\hline\hline \end{tabular} 
\begin{tabular} {| l l || c |c | c | c |} 
\multicolumn{6}{l}{Panel B: Results for topics with $tr\leq 5$} \\
\hline \hline 
{\em Model} & {\em Profile} & { F-Score } & { Accuracy } & { Precision } & { Recall } \\ \hline \hline{\em C4.5 } & {\em Tokens } & { 0.39 } & { 0.74 } & { 0.35 } & { 0.5 } \\
{\em } & {\em Wiki } & { 0.41 } & { 0.78 } & { 0.40 } & { 0.48 } \\
{\em } & {\em BTO Wiki } & { 0.41 } & { 0.77 } & { 0.40 } & { 0.49 } \\ \hline
{\em LibSVM } & {\em Tokens } & { 0.40 } & { 0.85 } & { 0.56 } & { 0.39 } \\
{\em } & {\em Wiki } & { 0.42 } & { 0.85 } & { 0.58 } & { 0.40 } \\
{\em } & {\em BTO Wiki } & { 0.44 } & { 0.84 } & { 0.55 } & { 0.43 } \\ \hline
{\em Wiki-SR } & {\em } & { 0.53 } & { 0.84 } & { 0.56 } & { 0.58 } \\
 \hline \hline \end{tabular} 
 \begin{tabular} {| l l || c |c | c | c |}
\multicolumn{6}{l}{Panel C: Results for topics with $tr >5$} \\
\hline \hline 
{\em Model} & {\em Profile} & { F-Score } & { Accuracy } & { Precision } & { Recall } \\ \hline \hline
{\em C4.5 } & {\em Tokens } & { 0.18 } & { 0.87 } & { 0.20 } & { 0.25 } \\
{\em } & {\em Wiki } & { 0.22 } & { 0.89 } & { 0.29 } & { 0.26 } \\
{\em } & {\em BTO Wiki } & { 0.20 } & { 0.89 } & { 0.25 } & { 0.23 } \\ \hline
{\em LibSVM } & {\em Tokens } & { 0.05 } & { 0.93 } & { 0.60 } & { 0.03 } \\
{\em } & {\em Wiki } & { 0.08 } & { 0.93 } & { 0.54 } & { 0.06 } \\
{\em } & {\em BTO Wiki } & { 0.09 } & { 0.92 } & { 0.53 } & { 0.07 } \\ \hline
{\em Wiki-SR} & {\em } & { 0.36 } & { 0.86 } & { 0.36 } & { 0.5 } \\
\hline \hline \end{tabular} 
\end{scriptsize}

\end{table}

First of all, a general comparison of the models suggests that the Wiki-SR heuristic achieves consistently better results than the benchmark algorithms in terms of F-score. See Figure~\ref{fig:comparisons} for F-score and Recall boxplots computed using all 100 topics. When searching for causes, it appears that the performance differences are largely explained by the recall levels. Whereas the differences in accuracies and precisions are relatively small, as observed from Table~\ref{tab:comparison}, the heuristic Wiki-SR achieves considerably better results in terms of recall. Interestingly, when considering a division of topics based on the proportion of irrelevant and relevant documents in the training sample, we find that the heuristic has faired considerably better than its benchmarks on highly unbalanced topics; see Panels B and C in Table~\ref{tab:comparison}. This observation is possibly explained by that a rule-based model such as Wiki-SR is less sensitive to the quality of the training sample than for example SVM-classifiers. It is also known that imbalance between positive and negative examples in training sample can have an adverse effect on traditional classifiers.

Finally, to investigate the effect of document model given to the benchmark classifiers, both SVM and C4.5 models were built using three alternative profiles with different levels of concept information. A quick comparison reveals that wikification slightly improves results for all topics as measured by F-Score. Especially, when more unbalanced topics are considered. However, the case of BTO concepts shows mixed evidence.

\section{Conclusions}\label{sec:conclusion}

In this paper, we have presented a new document filtering framework, Wiki-SR, where Wikipedia's extensive domain-knowledge is utilised to produce effective semantic classification rules. An empirical experiment based on Reuters RCV1 corpus and TREC-11 topic statements revealed that the use of semantic concept-relatedness information along with a suitable document model have a considerable combined effect on classification performance. The results suggest that although there are some benefits already in the use of a concept-based representation of document's contents, the profile is not truly effective unless there is also knowledge about relationships between different concepts. For this purpose, the use of Wikipedia as a source of domain knowledge is ideal due to its incredibly dense link-structure and broad scope. 

In the future work, we investigate how machine-learning can be used to complement our Wikipedia-based approach to determining document-concept relatedness. In particular, we assume that the techniques used in multi-task learning could prove to be very beneficial in this respect. As another direction for further development, we are examining how the boolean rules used in Wiki-SR can be better extracted automatically from text in natural language form. Especially, we are interested in considering techniques, where the rule structures can be learned without the use of explicit topic definitions. One of such directions is examined in the forthcoming paper Malo et al.~\cite{malo10b}, where a modified GP-algorithm is developed to learn Wikipedia-based queries using only sample documents supplied by the user.

\section*{Acknowledgment}

The authors would like to thank Emil Aaltonen Foundation and Finnish Cultural Foundation for their support.

\bibliographystyle{IEEEtran}
\bibliography{filtering}

% Generated by IEEEtran.bst, version: 1.13 (2008/09/30)
\begin{thebibliography}{10}
\providecommand{\url}[1]{#1}
\csname url@samestyle\endcsname
\providecommand{\newblock}{\relax}
\providecommand{\bibinfo}[2]{#2}
\providecommand{\BIBentrySTDinterwordspacing}{\spaceskip=0pt\relax}
\providecommand{\BIBentryALTinterwordstretchfactor}{4}
\providecommand{\BIBentryALTinterwordspacing}{\spaceskip=\fontdimen2\font plus
\BIBentryALTinterwordstretchfactor\fontdimen3\font minus
  \fontdimen4\font\relax}
\providecommand{\BIBforeignlanguage}[2]{{%
\expandafter\ifx\csname l@#1\endcsname\relax
\typeout{** WARNING: IEEEtran.bst: No hyphenation pattern has been}%
\typeout{** loaded for the language `#1'. Using the pattern for}%
\typeout{** the default language instead.}%
\else
\language=\csname l@#1\endcsname
\fi
#2}}
\providecommand{\BIBdecl}{\relax}
\BIBdecl

\bibitem{medelyan09}
O.~Medelyan, D.~Milne, C.~Legg, and I.~Witten, ``Mining meaning from
  \text{Wikipedia},'' \emph{International Journal of Human-Computer Studies},
  vol.~67, pp. 716--754, 2009.

\bibitem{milne07}
D.~Milne, ``Computing semantic relatedness using \text{Wikipedia} link
  structure,'' in \emph{Proceedings of the New Zealand Computer Science
  Research Student Conference}, 2007.

\bibitem{milne08}
D.~Milne and I.~Witten, ``Learning to link with \text{Wikipedia},'' in
  \emph{Proc. CIKM}, 2008.

\bibitem{gabrilovich06}
E.~Gabrilovich and S.~Markovitch, ``Overcoming the brittleness bottleneck using
  \text{Wikipedia},'' in \emph{Proc. National Conference on Artificial
  Intelligence}, Boston, MA, 2006.

\bibitem{gabrilovich07}
------, ``Computing semantic relatedness using \text{Wikipedia}-based explicit
  semantic analysis,'' in \emph{Proc. IJCAI-07}, 2007.

\bibitem{medelyan08a}
O.~Medelyan, I.~Witten, and D.~Milne, ``Topic indexing with \text{Wikipedia},''
  in \emph{Proceedings of the AAAI 2008 Workshop on Wikipedia and Artificial
  Intelligence (WIKIAI 2008)}, 2008.

\bibitem{medelyan08b}
O.~Medelyan and D.~Milne, ``Augmenting domain-specific thesauri with knowledge
  from \text{Wikipedia},'' in \emph{Proceedings of the New Zealand Computer
  Science Research Student Conference}, 2008.

\bibitem{mihalcea07}
R.~Mihalcea and A.~Csomai, ``Wikify!: linking documents to encyclopedic
  knowledge,'' in \emph{Proc. CIKM}, 2007, pp. 233--242.

\bibitem{strube06}
M.~Strube and S.~Ponzetto, ``{WikiRelate! Computing semantic relatedness using
  Wikipedia},'' in \emph{Proceedings of the 21nd AAAI conference on artificial
  intelligence}, 2006.

\bibitem{gregorowicz06}
A.~Gregorowicz and M.~Kramer, ``{Mining a large-scale term-concept network from
  Wikipedia},'' Mitre Corporation, Tech. Rep., 2006.

\bibitem{milne07b}
D.~Milne, I.~Witten, and D.~Nichols, ``A knowledge-based search engine powered
  by {Wikipedia}.'' in \emph{Proceedings of the 16th ACM Conference on
  Information and Knowledge Management CIKM'07}, 2007, pp. 445--454.

\bibitem{li07}
Y.~Li, R.~Luk, E.~Ho, and K.~Chung, ``{Improving weak ad-hoc queries using
  Wikipedia as external corpus},'' in \emph{Proceedings of the 30th Annual
  International ACM SIGIR Conference on Research and Development in Information
  Retrieval}.\hskip 1em plus 0.5em minus 0.4em\relax ACM Press, Amsterdam,
  2007, pp. 797--798.

\bibitem{egozi08}
O.~Egozi, E.~Gabrilovich, and S.~Markovitch, ``Concept-based feature generation
  and selection for information retrieval,'' in \emph{Proceedings of the 23rd
  AAAI Conference on Artificial Intelligence (AAAI-08)}, 2008.

\bibitem{wang08a}
P.~Wang, C.~Domeniconi, and J.~Hu, ``{Cross-domain Text Classification using
  Wikipedia},'' \emph{IEEE Intelligent Informatics Bulletin}, vol.~9, pp.
  5--17, 2008.

\bibitem{wang08b}
P.~Wang and C.~Domeniconi, ``{Building Semantic Kernels for Text Classification
  using Wikipedia},'' in \emph{KDD'08}, 2008.

\bibitem{finkel05}
J.~Finkel, T.~Grenader, and C.~Manning, ``Incorporating non-local information
  into information extraction systems by \text{Gibbs} sampling,'' in
  \emph{Proceedings of the 43nd Annual Meeting of the Association for
  Computational Linguistics (ACL)}, 2005, pp. 363--370.

\bibitem{malo10}
P.~Malo and P.~Siitari, ``{A Context-aware Approach to User Profiling with
  Interactive Preference Learning},'' \emph{Aalto University working paper
  W-482}, 2010.

\bibitem{suchanek08}
F.~Suchanek, G.~Kasneci, and G.~Weikum, ``{YAGO: A Large Ontology from
  Wikipedia and WordNet},'' \emph{Elsevier Journal of Web Semantics}, 2008.

\bibitem{cilibrasi07}
R.~Cilibrasi and P.~Vitanyi, ``The \text{Google} similarity distance,''
  \emph{IEEE Transactions on Knowledge and Data Engineering}, vol.~19, pp.
  370--383, 2007.

\bibitem{malo10b}
P.~Malo, P.~Siitari, and A.~Sinha, ``{Automated Query Learning with Wikipedia
  and Genetic Programming},'' \emph{Unpublished manuscript}, 2010.

\bibitem{smith97}
M.~Smith and M.~Smith, ``{The use of genetic programming to build boolean
  queries for text retrieval through relevance feedback},'' \emph{Journal of
  Information Science}, vol.~23, no.~6, pp. 423--431, 1997.

\bibitem{chenshanka98}
H.~Chen, G.~Shankaranarayanan, L.~She, and A.~Iyer, ``{A machine learning
  approach to inductive query by example: An experiment using relevance
  feedback, ID3, genetic algorithms, and simulated annealing},'' \emph{Journal
  of the American Society for Information Science}, vol.~49, no.~8, pp.
  693--705, 1998.

\bibitem{koza92}
J.~Koza, \emph{Genetic programming: On the programming of computers by means of
  natural selection}.\hskip 1em plus 0.5em minus 0.4em\relax MIT Press, 1992.

\bibitem{bontcheva04}
K.~Bontcheva, V.~Tablan, D.~Maynard, and H.~Cunningham, ``{Evolving GATE to
  Meet New Challenges in Language Engineering},'' \emph{Natural Language
  Engineering}, vol.~10, pp. 349---373, 2004.

\bibitem{milne09}
D.~Milne and I.~Witten, ``An open-source toolkit for mining \text{Wikipedia},''
  \emph{Unpublished manuscript}, 2009.

\bibitem{hall09}
M.~Hall, E.~Frank, G.~Holmes, B.~Pfahringer, P.~Reutemann, and I.~Witten,
  ``{The WEKA Data Mining Software: An Update},'' \emph{SIGKDD Explorations},
  vol.~11, pp. 10--18, 2009.

\bibitem{abeel09}
T.~Abeel, Y.~de~Peer, and Y.~Saeys, ``{Java-ML: A Machine Learning Library},''
  \emph{Journal of Machine Learning Research}, vol.~10, pp. 931--934, 2009.

\end{thebibliography}

\end{document}